\documentclass[a4paper,11pt]{article}
\pdfoutput=1 

\usepackage{jinstpub} 

\usepackage{lineno}
\usepackage{siunitx}
\usepackage{textcomp}
\usepackage{subcaption}
\DeclareSIUnit\bar{bar}
\usepackage{multirow}

\title{Qualification of piezo-electric actuators for the MADMAX booster system at cryogenic temperatures and high magnetic fields}

\author[a]{E.~Garutti,}
\author[d]{H.~Janssen,}
\author[b,1]{D.~Kreikemeyer-Lorenzo,\note{Corresponding author.}}
\author[a,2]{C.~Krieger,\note{Corresponding author.}}
\author[c]{A.~Lindner,}
\author[b]{B.~Majorovits,}
\author[c]{J.~Schaffran,}
\author[d]{B.~van Bree}

\affiliation[a]{Institut für Experimentalphysik, Universität Hamburg,\\Luruper Chaussee 149, 22761 Hamburg, Germany}
\affiliation[b]{Max-Planck-Institut für Physik,\\Föhringer Ring 6, 80805 Munich, Germany}
\affiliation[c]{Deutsches Elektronen-Synchrotron DESY,\\Notkestr. 85, 22607 Hamburg, Germany}
\affiliation[d]{JPE, Aziëlaan 12, 6199 AG Maastricht-Airport, The Netherlands}

\emailAdd{christoph.krieger@uni-hamburg.de}
\emailAdd{dkreike@mpp.mpg.de}

\abstract{We report on the qualification of a piezo-based linear stage for the manipulation of positions of dielectric discs in the booster of the MADMAX axion dark matter search experiment. A first demonstrator of the piezo drives, specifically developed for MADMAX, was tested at room temperature as well as at cryogenic temperatures down to \SI{4.5}{\kelvin} and inside strong magnetic fields up to \SI{5.3}{\tesla}. These qualification measurements prove that the piezo-based linear stage is suited for MADMAX and fulfills the requirements.}


\begin{document}
\maketitle
\flushbottom

\section{Introduction}
The MAgnetized Disc and Mirror Axion eXperiment (MADMAX)~\cite{brun2019} is a project to search for dark matter axions in the mass range around \SI{100}{\micro\eV}. This mass range, which is particularly difficult to access experimentally, is favored in the scenario where the breaking of the Peccei-Quinn symmetry happens after an early cosmic inflation~\cite{buschmann2022}. To explore this part of the axion parameter space, MADMAX utilizes the dielectric haloscope approach~\cite{caldwell2017}. The axion to photon conversion occurs inside a strong magnetic field at the boundaries between media with different dielectric constant $\varepsilon$.

The core element of MADMAX is the so-called \textit{booster}, an arrangement of dielectric discs of up to \SI{1.25}{\metre} diameter in front of a metallic mirror. The discs will be made out of a material with high $\varepsilon$ and low dielectric losses (low $\tan\delta$), like lanthanum aluminate or sapphire. The mirror directs the power generated in the booster system uni-directionally towards a receiver at one end of the system. By arranging the dielectric discs in a proper way, the very weak emitted power (from dark matter axions) of the order of $10^{-27}$ \SI{}{\watt} can be boosted to a detectable level. The frequency dependent ratio between power emitted by the booster to power emitted by the mirror only is called boost factor. By re-arranging the distance between discs, the axion mass to which the system is sensitive, as well as the bandwidth of the boost factor, can be tuned. The booster will be housed inside a cryostat and will be cooled to liquid helium temperature in order to increase sensitivity of the receiver by minimising the thermal noise of the detector and booster. The entire set-up will be placed inside a $\sim\SI{9}{\tesla}$ dipole magnet to be located in the iron yoke of the former H1 experiment at DESY in Hamburg (Germany). The power emitted from the booster in form of microwave radiation is then focused by an elliptical mirror onto a horn antenna connected to a low-noise receiver system using a heterodyne detection scheme.

To tune the range in which the experiment is sensitive, it is necessary to re-arrange the distance between discs. For this purpose, a high-precision positioning system is required to move the large discs (approximate weight of a \SI{1.25}{\metre} diameter lanthanum aluminate disc is \SI{6}{\kilo\gram}). The positioning system has to withstand cryogenic temperatures (around \SI{4}{\kelvin}), high magnetic fields (\SI{9}{\tesla}) and has to ensure high precision (better than \SI{10}{\micro\metre})  along a long stroke in the order of typically many centimetres up to, in extreme scenarios, a metre. Additionally, the positioning system needs to be able to operate in a gaseous helium atmosphere, which is necessary to ensure an efficient cooldown and a good thermalisation of the dielectric discs. A concept for the positioning system, using piezo actuators based on the stick-slip principle, was developed by the MADMAX collaboration in cooperation with the company JPE: three actuators are used to move a disc, thus featuring actuators with different orientations with respect to gravity and mechanical load. Table~\ref{tab:table1} lists the specifications for the piezo-based linear stage (piezo actuator plus carriage) and a summary of the performed qualification tests.  

\begin{table}[ht]
  \begin{center}
    \caption{Technical specifications of the MADMAX positioning system, and summary of conducted tests at room temperature (RT) and cryogenic temperatures (low T).}
    \label{tab:table1}
    \begin{tabular}{|l|c|c|c|} 
      \hline
      \textbf{Parameter} & \textbf{Requirement}& \textbf{Tested at RT} & \textbf{Tested at low T}\\
      \hline
      \multicolumn{4}{|c|}{General specifications} \\
      \hline
      Power consumption at rest & $<\SI{25}{\milli\watt}$ &  & \\
      \hline
      Power consumption during operation & $<\SI{40}{\milli\watt}$ &  & \\
      \hline
      Heat input through cabling & $<\SI{0.5}{\milli\watt}$ & & \\
      \hline
      Operation in magnetic field & \SIrange{9}{10}{\tesla} & not tested & \SI{5.3}{\tesla}\\
      \hline
      Operation at cryogenic temperatures & \SI{4.2}{\kelvin}& N/A & \SI{4.2}{\kelvin}\\
      \hline
      Operation in vacuum & \SI{e-6}{\milli\bar} & ok & ok \\
      \hline
      Operation in He gas atmosphere & \SI{e-1}{\milli\bar} & N/A & \SI{e-1}{\milli\bar}\\
      \hline
      \multicolumn{4}{|c|}{Motor requirements} \\
      \hline
      Travel range & \SI{1000}{\milli\metre}& \SI{130}{\milli\metre}$^\ast$ & \SI{130}{\milli\metre}$^\ast$\\
      \hline
      Stage velocity at \SI{4.2}{\kelvin} & \SI{100}{\micro\metre\per\second}& N/A & \SIrange{60}{100}{\micro\metre\per\second}\\
      \hline
      Stepsize & $<\SI{10}{\micro\metre}$ at \SI{4.2}{\kelvin} & \SI{12}{\micro\metre} &  \SI{2}{\micro\metre}\\
      \hline
      Load per actuator & \SI{2}{\kilo\gram} & \SI{2}{\kilo\gram} & \SI{2}{\kilo\gram}\\
      \hline
      Different actuator orientations & see figure~\ref{fig:rt-setup} & ok &  ok\\
      \hline
      Lifetime & \SI{10}{\metre} & & $>\SI{13}{\metre}^{\ast\ast}$ \\
      \hline
      \multicolumn{2}{l}{$^\ast$ maximum travel range in our set-up} \\
      \multicolumn{2}{l}{$^\ast$$^\ast$ factory acceptance test} \\
    \end{tabular}
  \end{center}
\end{table}

\section{The stick-slip piezoelectric linear stage for MADMAX}
\begin{figure}
    \centering
    \subcaptionbox{\label{fig:jpe-mm1-carriage}}{\includegraphics[height=4.2cm]{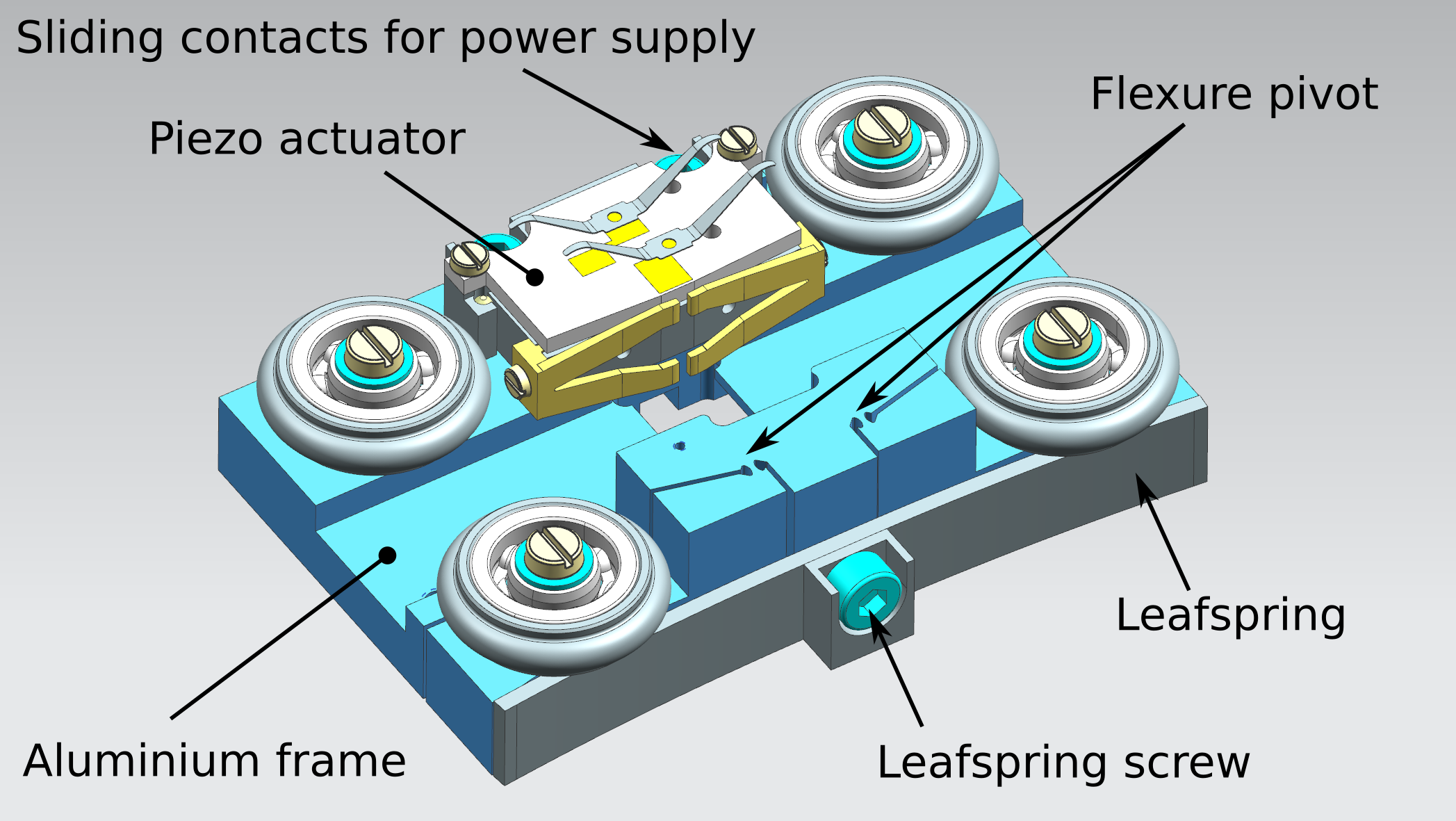}}
    \subcaptionbox{\label{fig:jpe-mm1-rail}}{\includegraphics[height=4.2cm]{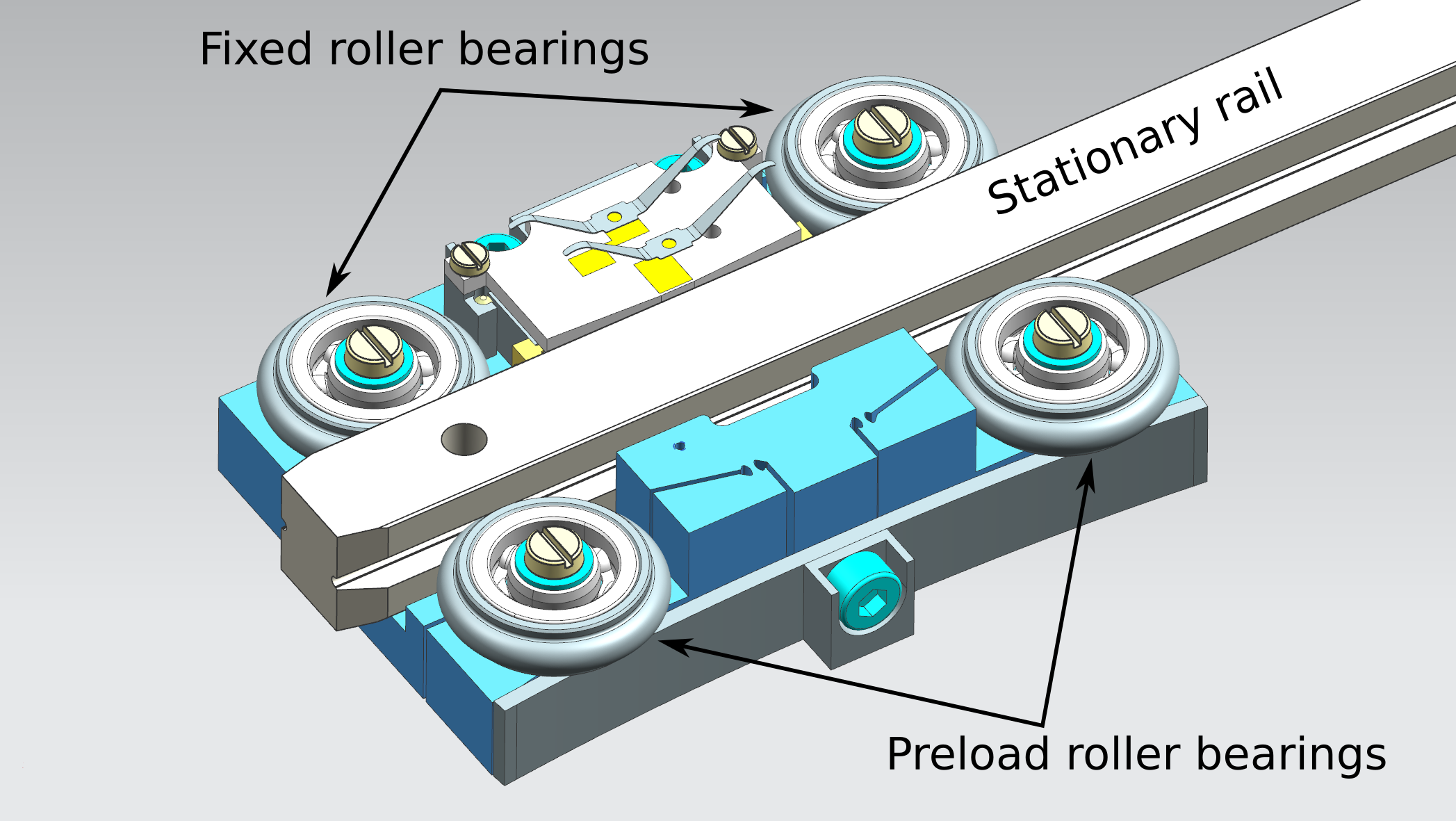}}
    \caption{The MADMAX linear stage consists of a carriage with piezo actuator and roller bearings (\subref{fig:jpe-mm1-carriage}) interfacing with the ceramic rail (\subref{fig:jpe-mm1-rail}).}
    \label{fig:jpe-mm1-labeled}
\end{figure}

\begin{figure}
    \centering
    \subcaptionbox{\label{fig:jpe-mm1-concept}}{\includegraphics[height=3cm]{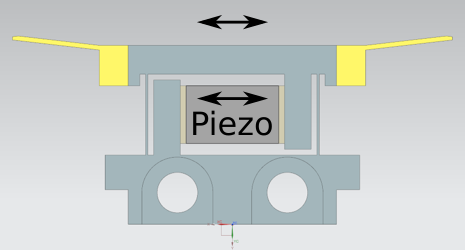}}
    \subcaptionbox{\label{fig:jpe-mm1-pic1}}{\includegraphics[height=3cm]{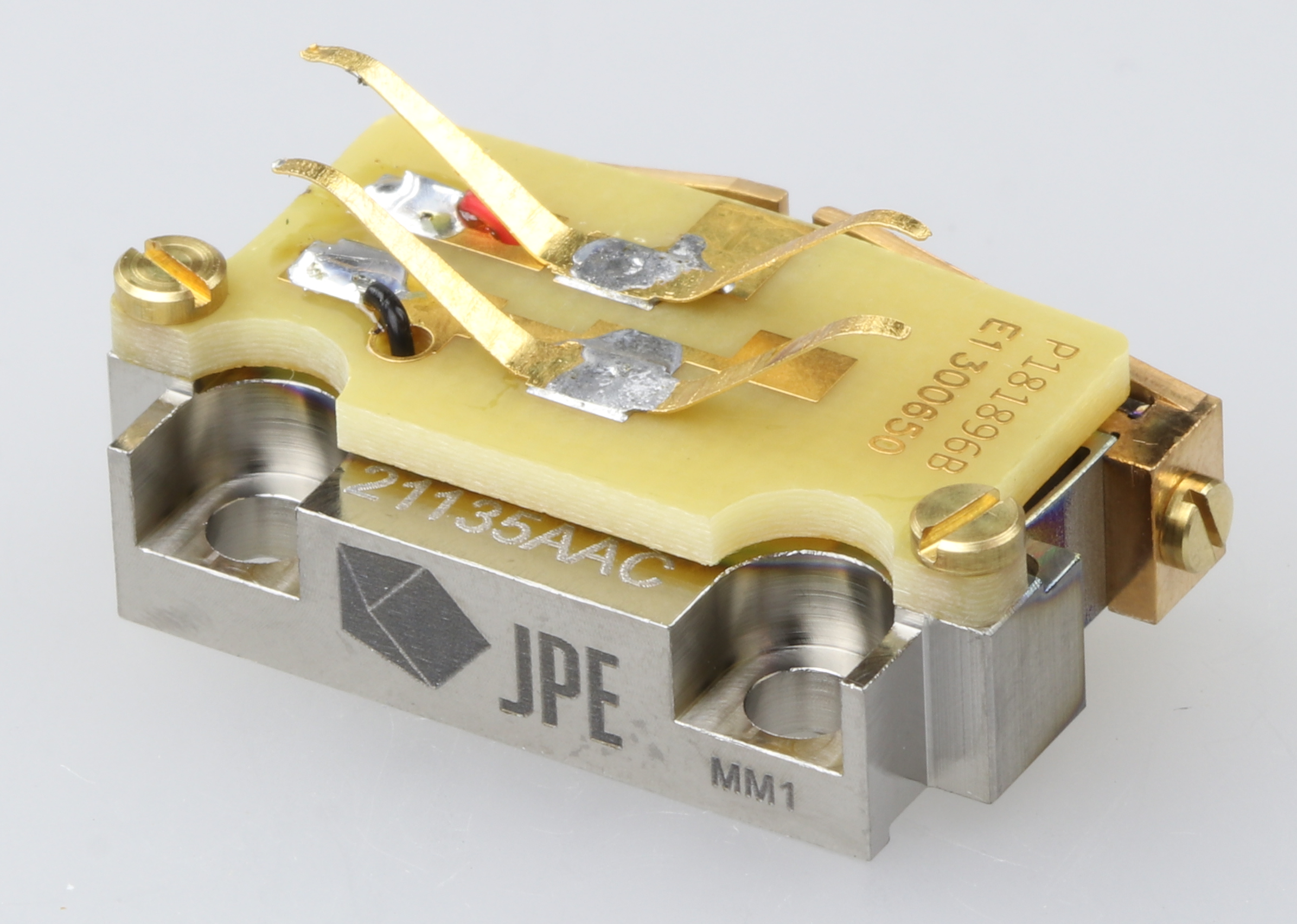}}
    \subcaptionbox{\label{fig:jpe-mm1-pic2}}{\includegraphics[height=3cm]{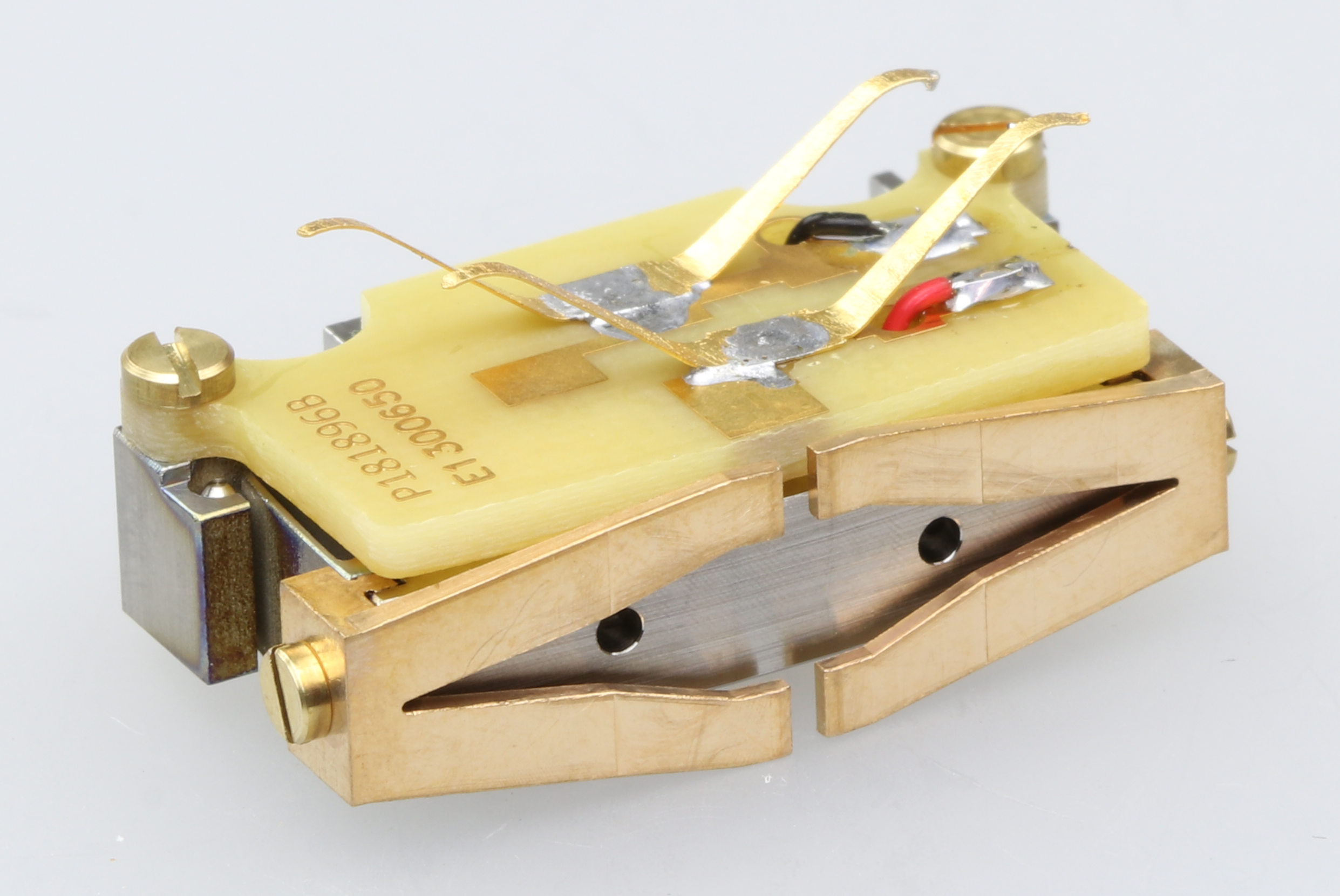}}
    \caption{Piezo actuator concept (\subref{fig:jpe-mm1-concept}) and realized prototype (\subref{fig:jpe-mm1-pic1} and \subref{fig:jpe-mm1-pic2}). In (\subref{fig:jpe-mm1-concept}) the wings (yellow) of the actuator are pressed against the stationary rail not shown here. Compare also to the component labeled "Piezo actuator" in figure~\ref{fig:jpe-mm1-labeled}.}
    \label{fig:jpe-mm1}
\end{figure}

The disc spacing in the booster must be adjustable and therefore each disc is connected to three carriages, in a \num{120} degrees distribution along the disc circumference. Each carriage runs over a stationary ceramic rail, as shown in figure~\ref{fig:jpe-mm1-labeled}. A carriage interfaces to its rail via four ceramic roller bearings, two fixed and two mounted on a flexure pivot, allowing them to be spring preloaded to maintain play free and smooth operation, to compensate for significant dimensional changes between assembly at room temperature and operation at cryogenic temperature. The preload force is applied by simply fixing a single screw, no adjustment is needed. The screw will deform a leafspring to a predetermined value, set by geometry in the aluminum carriage frame, and the symmetrical layout guarantees that both bearings are preloaded with the same force.
To drive a carriage over the rail it is fitted with a bespoke JPE designed MADMAX actuator (referred as piezo actuator in figure~\ref{fig:jpe-mm1-labeled}). This piezo electric actuator operates by the so-called stick-slip motion and is depicted in figure~\ref{fig:jpe-mm1}. When applying the above-mentioned bearing preload, the wings of the actuator (yellow parts at the top in figure~\ref{fig:jpe-mm1-concept}) are pressed at the same time against the stationary rail with a predefined force, thereby locking this part of the actuator to the rail in the direction of motion by friction. For slow piezo movement the inertia of the carriage with the connected disc will not be sufficient to break this friction, and the carriage will move with respect to the wings and therefore also with respect to the rail, but a rapid piezo return-motion will cause the friction to break, and the carriage will remain stationary. Effectively, a micron level step is made and repetitive operation of the actuator will result in macroscopic motion, up to the required meter level. 
An additional challenge is the power supply to the actuator, which traditionally would require wires to travel with the carriage over a meter distance. An elegant solution has been found in the use of sliding contacts on the piezo actuator which contact a stationary power rail in the booster, thus eliminating the need for wires. 

\section{Qualification at room temperature}

To qualify the demonstrator of the MADMAX piezolectric linear stage at room temperature, the displacement was measured as function of time for different settings of the relevant actuator parameters (relative stepsize and frequency of the steps), different mechanical loads and in different orientations.
\subsection{Set-up}

\begin{figure}
    \centering
    \includegraphics[width=\textwidth]{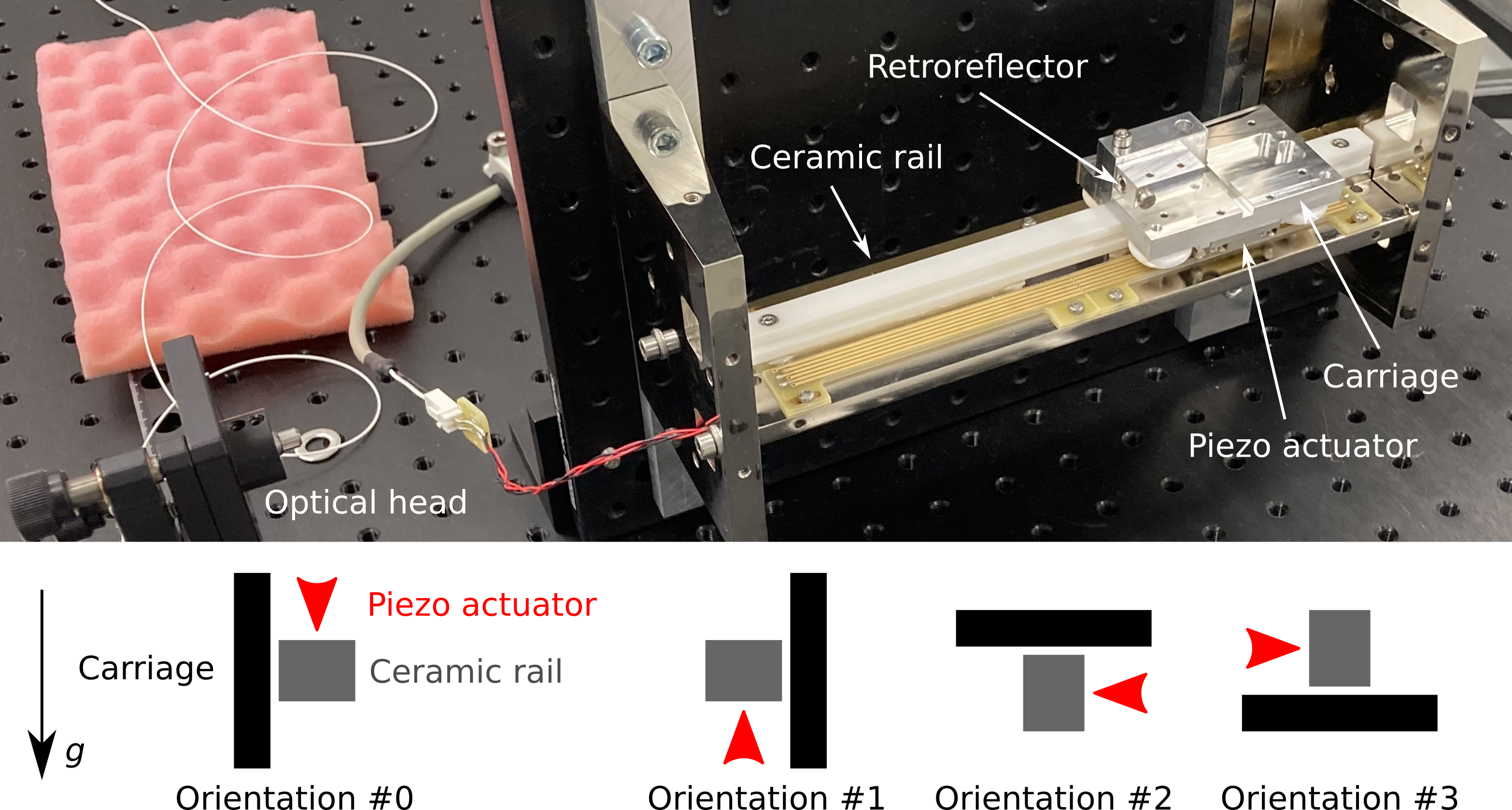}
    \caption{Picture showing the room temperature set-up (without the additional weight block) and main parts labeled. Sketches below the picture indicate the nomenclature of the different orientations used throughout this text. The picture shows the set-up in orientation \#2.}
    \label{fig:rt-setup}
\end{figure}

The set-up of the demonstrator is shown in figure~\ref{fig:rt-setup}, which features a single actuator assembly of which in MADMAX three will be used per disc. The set-up can be placed in the four different orientations (\#0, \#1, \#2, \#3) indicated in the lower part of figure~\ref{fig:rt-setup}. In all orientations the movement of the stage is in the horizontal plane, as foreseen for the MADMAX booster. A retroreflector was mounted on the carriage (or on the weight block) in order to measure its displacement along the ceramic rail. It was aligned with the optical head of a laser interferometer system (SmarAct PicoScale with C03 sensor head and environmental module) which allows to measure the stage displacement with a frequency of approximate \SI{1}{\kilo\hertz} and a resolution of better than \SI{100}{\nano\metre}.

A weight block (made of two parts) can be mounted onto the carriage to test the system with different mechanical loads and to simulate the weight of the disc: no weight (\SI{0}{\kilo\gram}), \SI{1}{\kilo\gram} (half weight block), and \SI{2}{\kilo\gram} (full weight block). The \SI{2}{\kilo\gram} weight configuration resembles the situation in the booster where a single \SI{6}{\kilo\gram} disc is moved by a total of three motors.

Communication with the linear stage was realized via a dedicated CADM2 controller by JPE (which was customized for the operation of the demonstrator). Input parameters to the controller are the relative stepsize (RSS) and step frequency. The RSS can be varied from \SIrange{0}{100}{\percent} in \SI{1}{\percent} increments and affects the maximum voltage difference applied to the piezo actuator during a stroke and by this the achieved displacement in one step of the stick-slip motion (stepsize). The step frequency, adjustable from \SIrange{0}{50}{\hertz} in \SI{1}{\hertz} increments, determines the number of steps performed per second. Multiplying stepsize (depending on RSS) and frequency gives the stage velocity.
\subsection{Measurements and results}

\begin{figure}
    \centering
    \subcaptionbox{\label{fig:jpe-steps-single}}{\includegraphics[height=5.4cm]{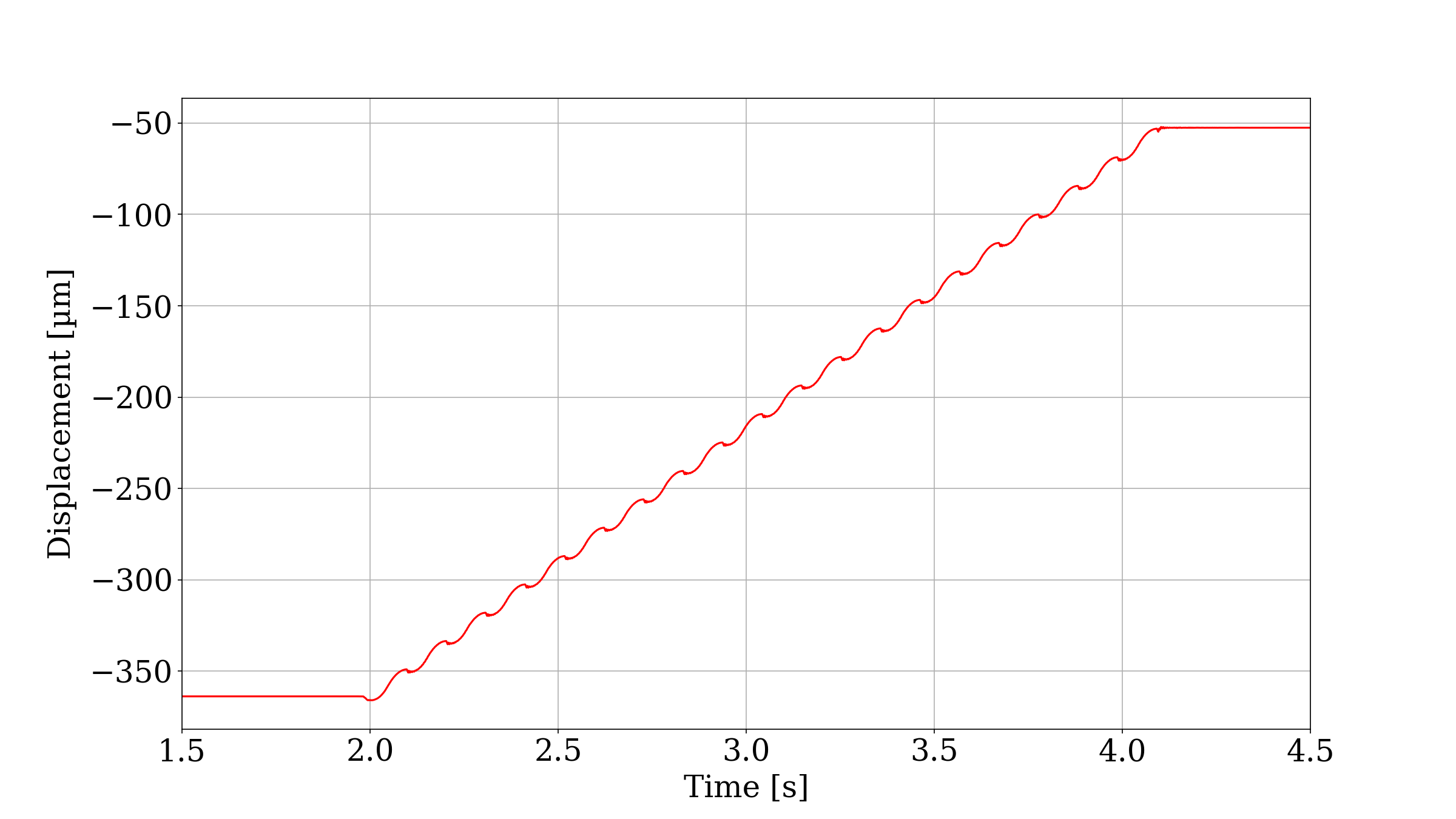}}
    \subcaptionbox{\label{fig:jpe-steps-histo}}{\includegraphics[height=5.4cm]{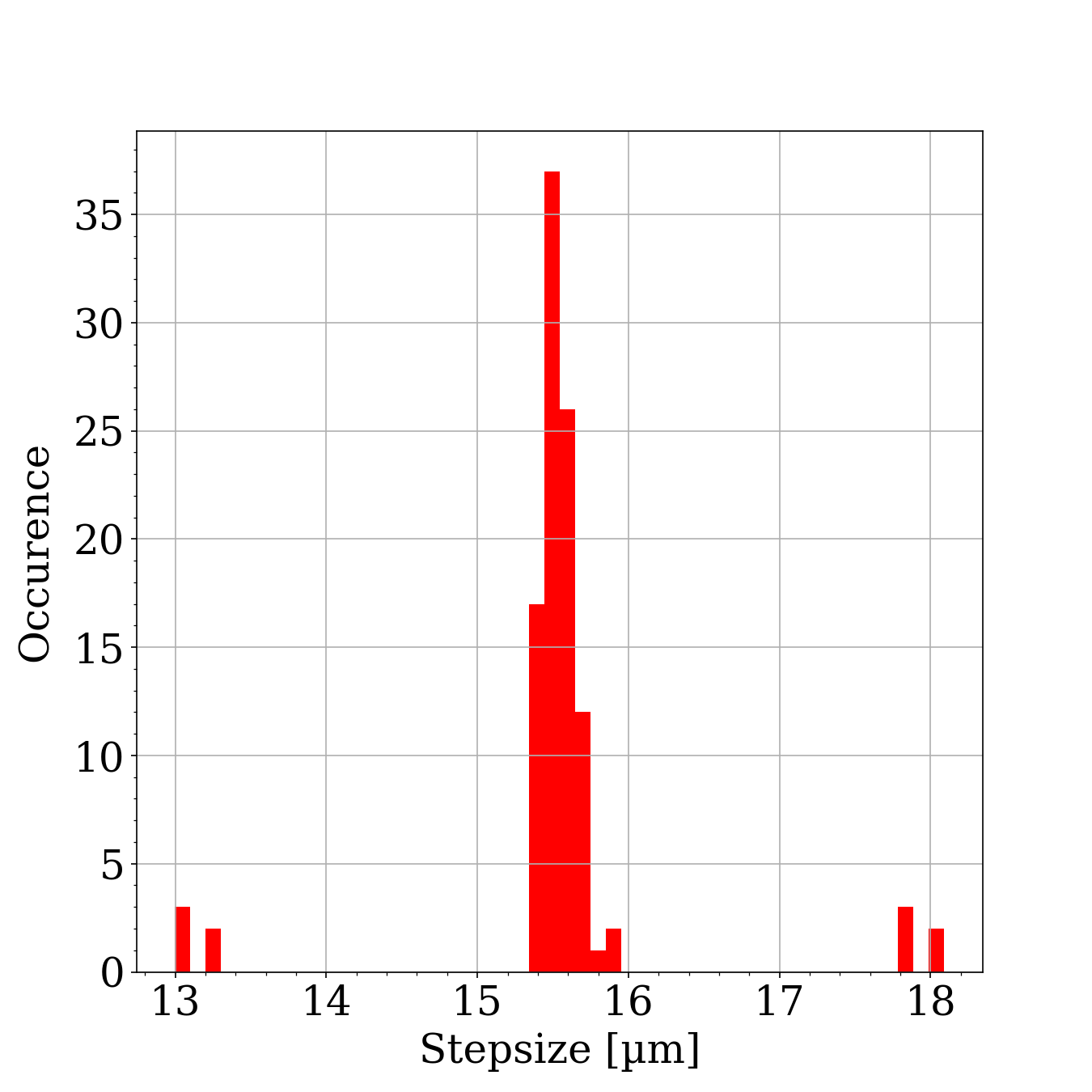}}
    \caption{Displacement of the piezo stage measured with the Picoscale laser interferometer during a forward movement of \num{20} steps with a frequency of \SI{10}{\hertz} and \SI{100}{\percent} stepsize~(\subref{fig:jpe-steps-single}), and histogram of the individual stepsizes extracted from five \num{20}-step forward movements~(\subref{fig:jpe-steps-histo}). The stage was operated in orientation \#0, with \SI{2}{\kilo\gram} mechanical load, at room temperature and without a magnetic field. In (\subref{fig:jpe-steps-single}) the individual steps and the stick-slip movement are clearly visible. In (\subref{fig:jpe-steps-histo}) a narrow distribution of the individual stepsizes around the mean stepsize is visible, due to technical reasons the first and last step in a movement are always different (one being shorter and the other longer than the steps in between) resulting in the two side peaks left and right of the main distribution. This does not affect the mean stepsize.}
    \label{fig:jpe-steps}
\end{figure}

To qualify the demonstrator at room temperature, overall stage velocity and average stepsize were measured as function of RSS and frequency for four different orientations and the three different weight configurations. To do so, the stage displacement was measured as function of time. Figure~\ref{fig:jpe-steps-single} shows the displacement with time for a sequence of \num{20} steps in forward direction with \SI{100}{\percent} relative stepsize and for \num{10} steps per second, in orientation \#0 and with \SI{2}{\kilo\gram} mechanical load. The individual steps are clearly visible as well as the small reverse movement after each forward step which is typical for the inertia-based stick-slip principle. The demonstrator was behaving as expected. The same behavior was observed in all orientations tested. Figure~\ref{fig:jpe-steps-histo} shows the distribution of the individual stepsizes extracted from five \num{20}-step movements: the two small peaks left and right of the mean stepsize are formed by the first and last steps of each \num{20}-step movement, as these are shorter/longer than the steps in between due to technical reasons. This does not affect the mean stepsize as the shortest and longest steps compensate each other. For the configuration used in figure~\ref{fig:jpe-steps}, the width of the distribution of the individual stepsizes is below \SI{100}{\nano\metre} (not taking into account the first and last step of each movement).

For the following measurements described in this section, the mean stepsize is extracted from five \num{20}-step movements. For the later measurements at cryogenic temperatures, no cryo-compatible laser interferometer was available at the time. The use of end switches in the set-up for cryogenic temperatures allows to compute the velocity of the stage from the travel time of a defined range. The velocity is therefore used as a figure of merit to allow easy comparison between room temperature and cryogenic measurements.

The velocity (and stepsize) of the stage was measured as a function of the step frequency (varied from \SIrange{10}{50}{\hertz} in \SI{10}{\hertz} increments) and the relative stepsize (varied from \SIrange{20}{100}{\percent} in four steps) for all four orientations and the three different weight configurations, forward and reverse. As to be expected, the velocity is a linear function of the step frequency, and independent of the orientation. Therefore, later measurements at cryogenic temperatures were carried out with a fixed frequency of \SI{50}{\hertz} and for only three of the four orientations.

\begin{figure}
    \centering
    \includegraphics[width=\textwidth]{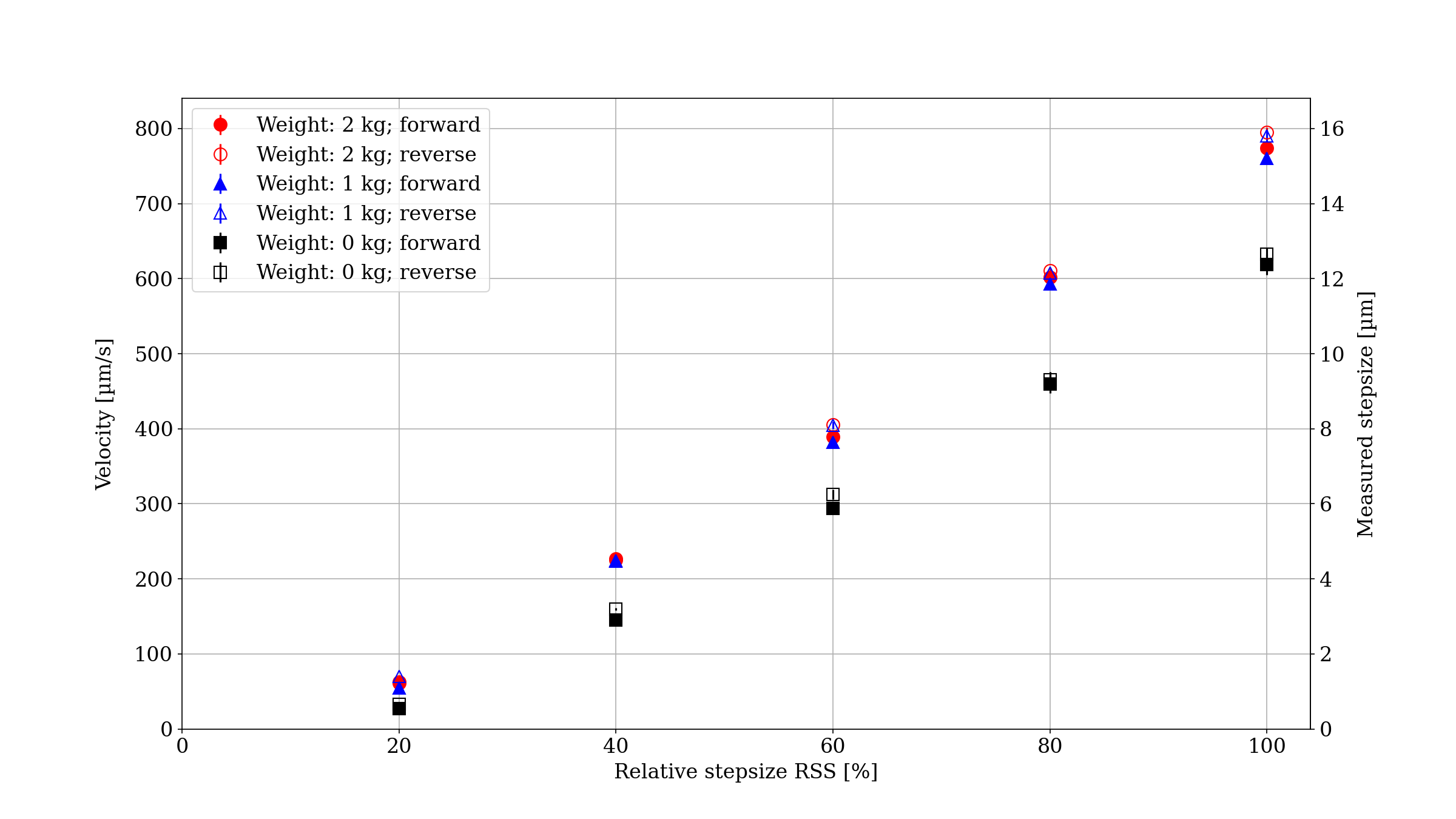}
    \caption{Linear stage velocity and stepsize measured with a laser interferometer as a function of the relative stepsize parameter for different loads mounted on the carriage of the linear stage (\SI{0}{\kilo\gram}, \SI{1}{\kilo\gram} and \SI{2}{\kilo\gram}). Measurements were performed with the actuator in orientation \#0, at room temperature, and without a magnetic field.}
    \label{fig:jpe-speed-stepsize-weight}
\end{figure}

Figure~\ref{fig:jpe-speed-stepsize-weight} shows the forward and reverse velocities (and stepsizes) as function of the relative stepsize for the three different weight configurations in orientation \#0 and for a step frequency of \SI{50}{\hertz}. For this setting it can be clearly seen that the determined velocity is proportional to the relative stepsize. This general behaviour was reproduced in the measurements with all other settings. As expected for an actuator using the stick-slip principle there is a small difference (typically in the order of less than a micron) between the forward and reverse stepsize. The stick-slip motion relies on the inertia of the load, hence a minimum mechanical load on the carriage is required for the system to work according to specifications. This can be seen in figure~\ref{fig:jpe-speed-stepsize-weight} as the velocity with \SI{0}{\kilo\gram} is actually lower (by about \SIrange{20}{50}{\percent} depending on the relative stepsize) than the velocity with an added mechanical load, while it hardly differs by less than a micron for \SI{1}{\kilo\gram} and \SI{2}{\kilo\gram}. Without sufficient inertia, the stage will move more in the opposite direction during the slip phase of the motion, resulting in a smaller stepsize.
The dependence of the velocity and stepsize on the relative stepsize parameter is linear with an offset, i.e., a minimum relative stepsize (or voltage difference in the applied signal) is needed for the stage to start moving.

\section{Qualification at cryogenic temperatures and high magnetic field}

The availability of a system that reliably actuates the discs with the needed precision at cryogenic temperature and at high magnetic field is one of the very challenging aspects of the MADMAX project. Especially, it is known that for these kind of piezoelectric actuators the stroke and thus the stage velocity decrease when going to lower (or even cryogenic) temperatures. In order to verify the functionality of the piezo-based linear stages, it is unavoidable to test them in a representative environment. Initial tests were performed in an available vacuum cryostat at a temperature of \SI{4.5}{\kelvin}. In a second step, the stage was also tested inside a \SI{5.3}{\tesla} dipole magnet, both at \SI{28}{\kelvin} in vacuum and at approximately \SI{5}{\kelvin} with helium exchange gas surrounding the piezo stage. For all these cryogenic tests the velocity was derived by measuring the time needed to move the stage between two end switches of the system. A summary of the different set-ups and performed measurements can be found in table~\ref{tab:table2}. 
It is important to mention that at cryogenic temperatures piezoelectric ceramics experience various changes in their properties, including a strongly reduced electrical capacitance, as well as a reduction of the strain coefficients (\cite{Simpson_1987, Locatelli_1988, Blackford_1992, Fouaidy_2007, Adhikari_2021}). These changes might influence the performance of the piezo actuators.

\begin{table}[htb]
  \begin{center}
    \caption{Summary of set-ups and measurements. At room temperature the velocity was measured with the laser interferometer over short movements and not full travel cycles. Measurements were done for the forward (f) as well as the reverse (r) velocity. Room temperature is abbreviated as RT, and gaseous helium as gHe.}
    \label{tab:table2}
    \small
    \begin{tabular}{|c|c|c|c|c|c|c|c|} 
      \hline
      \textbf{Stage set-up} & \textbf{Test facility} & \textbf{Orient.} & \textbf{$P [\SI{}{\milli\bar}]$}& \textbf{Temp}& \textbf{cycles} & \textbf{vel. (f)} & \textbf{vel. (r)}\\
      \hline
      JPE set-up & Lab & all & \num{e3} & RT & - & \SI{775(4)}{\micro\metre\per\second} & \SI{795(6)}{\micro\metre\per\second}\\
      \hline
      JPE set-up & Vac. Cryostat & \#3 & $<\num{e-5}$ & \SI{4.5}{\kelvin} & \num{10} & \SI{71(5)}{\micro\metre\per\second} & \SI{80(5)}{\micro\metre\per\second}\\
      \hline
      JPE set-up & Vac. Cryostat & \#1 & $<\num{e-5}$ & \SI{4.5}{\kelvin} & \num{8} & \SI{71(5)}{\micro\metre\per\second} & \SI{80(5)}{\micro\metre\per\second}\\
      \hline
      ALPS set-up & Vac. Cryostat & \#2 & $<\num{e-5}$ & \SI{4.5}{\kelvin} & \num{20} & \SI{116(5)}{\micro\metre\per\second} & \SI{116(5)}{\micro\metre\per\second}\\
      \hline
      ALPS set-up & ALPS-magnet & \#2 & $<\num{e-4}$ & \SI{28}{\kelvin} & \num{10} & \SI{126(10)}{\micro\metre\per\second} & \SI{142(6)}{\micro\metre\per\second}\\
      \hline
      ALPS set-up & ALPS-magnet & \#2 & \num{e-1} (gHe) & \SI{5}{\kelvin} & \num{6} & \SI{50(5)}{\micro\metre\per\second} & \SI{62(5)}{\micro\metre\per\second}\\
      \hline
    \end{tabular}
  \end{center}
\end{table}

\subsection{Set-ups for cryogenics}
Two different linear stage test set-ups were used for the cryogenic tests. For the tests at \SI{4.5}{\kelvin} (without a magnet), the original set-up delivered by JPE was slightly modified to mount it onto the cold plate of a Cryovac vacuum cryostat. This set-up was placed inside the vacuum volume of the Cryovac cryostat. The pressure reached was better than \SI{6e-6}{\milli\bar}. In order to ensure good heat transfer, the cold plate and the weight block of the linear stage were connected via a copper braid. Two end switches were installed to constrain the travel range of the stage to a defined distance. The controller of the linear stage was programmed to switch the moving direction when one of the end switches was activated. A \SI{2}{\kilo\gram} block was attached to the stage carriage to simulate the weight of a dielectric disc. The temperature of the system was monitored using three temperature sensors (LakeShore DT-670 silicon diodes). The sensors were attached to one side plate, to the copper strip (responsible for heat transfer between stage and cryostat) and to the weight block. The linear stage velocity, obtained from the measured travelling time, is used to characterize the performance of the stage under different conditions.

\begin{figure}
    \centering
    \includegraphics[width=\textwidth]{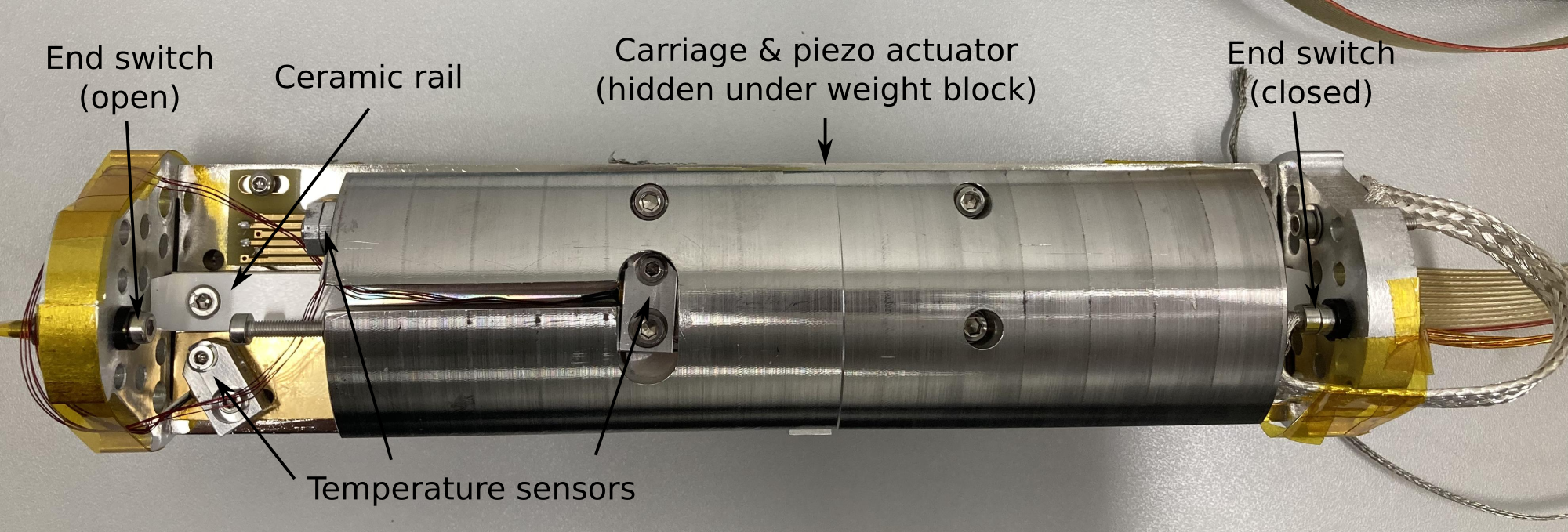}
    \caption{Demonstrator set-up modified to fit in the bore of the ALPS magnet. Below the \SI{1}{\kilo\gram} weight block, the unmodified carriage is running along the unaltered ceramic rail. The overall length of the set-up shown is about \SI{24}{\centi\metre}.} 
    \label{fig:alps-set-up}
\end{figure}

For the tests in the magnetic field in vacuum and in a gaseous environment, an ALPS-II dipole magnet~\cite{Albrecht_2021} at DESY (Deutsches Elektronen-Synchrotron) was used. In order to meet the limited space requirements of the \SI{55}{\milli\metre} diameter cold bore, modifications were carried out. The modified set-up including the modified test weight are shown in figure~\ref{fig:alps-set-up}. In this ALPS-magnet test set-up, smaller side plates were required, as well as a more compact stainless steel weight (approximately \SI{1}{\kilo\gram}). Three CERNOX temperature sensors were attached to the side of the weight, to the cooling copper strip and to the carriage (closer to the piezo actuator than in the previous configuration), see figure~\ref{fig:alps-set-up} for their position. The magnetic field is perpendicular to the direction of motion of the stage as it will be in the MADMAX booster.

\subsection{Qualification measurements at cryogenic temperatures}
\begin{figure}
  \includegraphics[width=\textwidth]{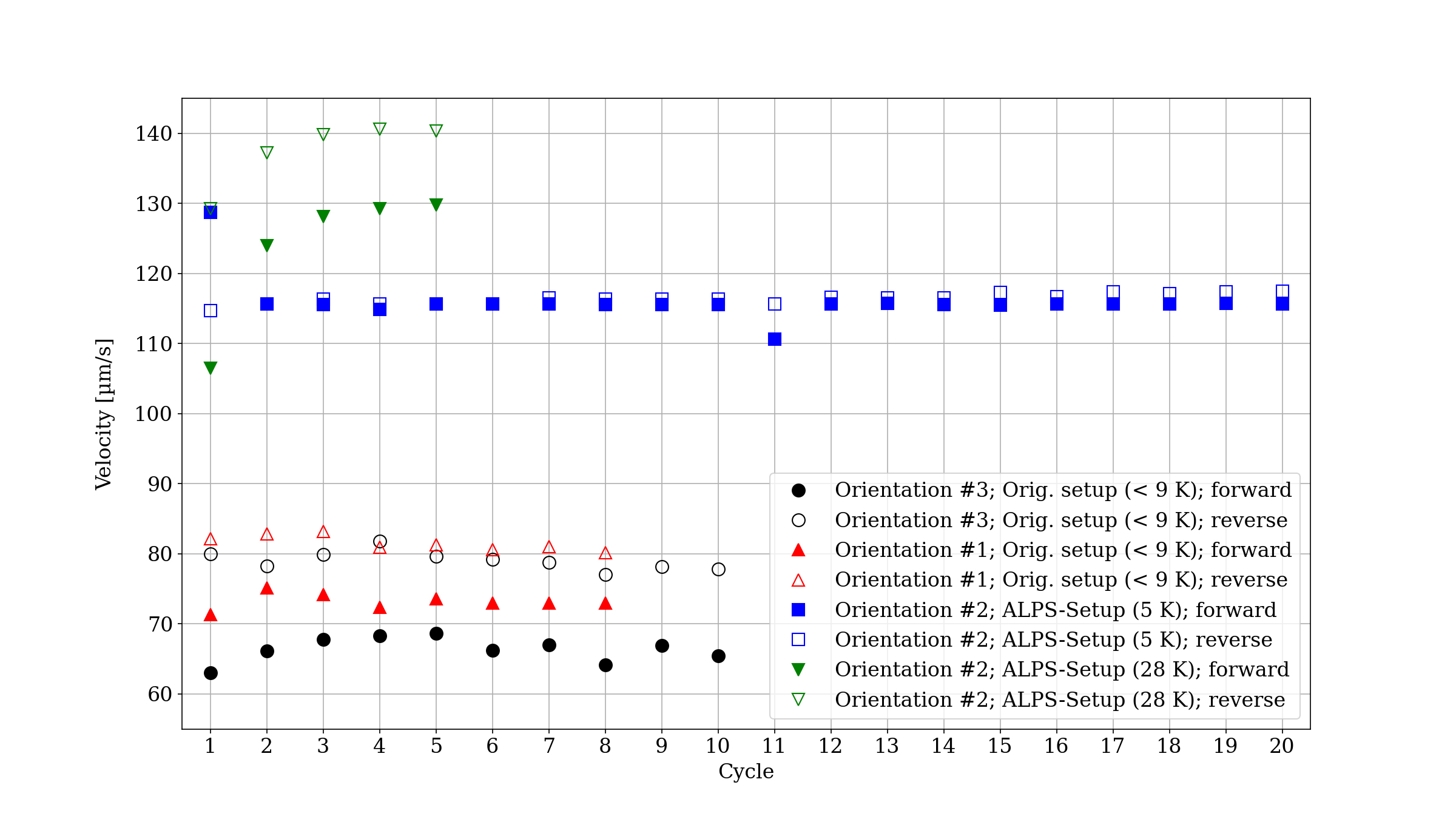}
  \caption{Velocity of the linear stage measured with the original set-up and the ALPS set-up at different cryogenic temperatures and in different orientations. All tests were carried out in a vacuum cryostat, except the measurement at \SI{28}{\kelvin}, performed inside the ALPS-magnet vacuum bore. The measurements were performed without a magnetic field. The maximum error is smaller than $\pm $\SI{5}{\micro\metre\per\second}, decreasing with decreasing stage velocity. Errorbars are not shown for the sake of simplicity. } 
  \label{fig:speed_2orient}
\end{figure}

Figure~\ref{fig:speed_2orient} shows the stage velocity derived from the test runs for three different actuator orientations (all following a horizontal movement). Three of the data sets (black, red and blue markers) were taken in a vacuum cryostat at temperatures below \SI{10}{\kelvin}, whereas the last one (green markers) was taken in the ALPS magnet at \SI{28}{\kelvin}. At the beginning of each run in the vacuum cryostat, the temperature of the stage was \SI{4.5}{\kelvin}, gradually increasing during operation to a maximum of \SI{9}{\kelvin}. The stage moved along a travel range of \SI{14}{\milli\metre} for \num{10} cycles in orientation \#3 (see figure~\ref{fig:rt-setup}, bottom) and along an \SI{18}{\milli\metre} range for \num{8} cycles at orientation \#1. A cycle is defined as the movement of the stage from the start to the end position and back to the start position.  Forward and reverse movement directions are plotted separately in order to visualize any direction dependencies. 

With the JPE set-up in the vacuum cryostat, the stage moved reliably in both orientations, with a slightly different velocity in orientation \#1 than in orientation \#3. A stage velocity of approximately \SI{71}{\micro\metre\per\second} in the forward direction and approximately \SI{80}{\micro\metre\per\second} in the reverse direction was derived. Before testing the stage in the presence of the magnetic field, the ALPS set-up was tested for \num{20} cycles in the vacuum cryostat (blue markers). The absolute stage velocity was significantly faster than with the original set-up (approximately \SI{116}{\micro\metre\per\second}). However, the hysteresis between the two movement directions was almost absent. The improvement in the velocity was likely caused by a change of the carriage wheels/bearings to an improved design which was implemented along with the general modification of the set-up for the tests in the magnet. The ALPS set-up was tested in the ALPS magnet after cool down, in vacuum, initially without a magnetic field (green markers). The minimum temperature achieved inside the magnet (in vacuum) was \SI{28}{\kelvin}. Due to the higher temperature, the stage was significantly faster: approximately \SI{124}{\micro\metre\per\second} in the forward direction and \SI{137}{\micro\metre\per\second} in the reverse direction.

\begin{figure}
  \includegraphics[width=\textwidth]{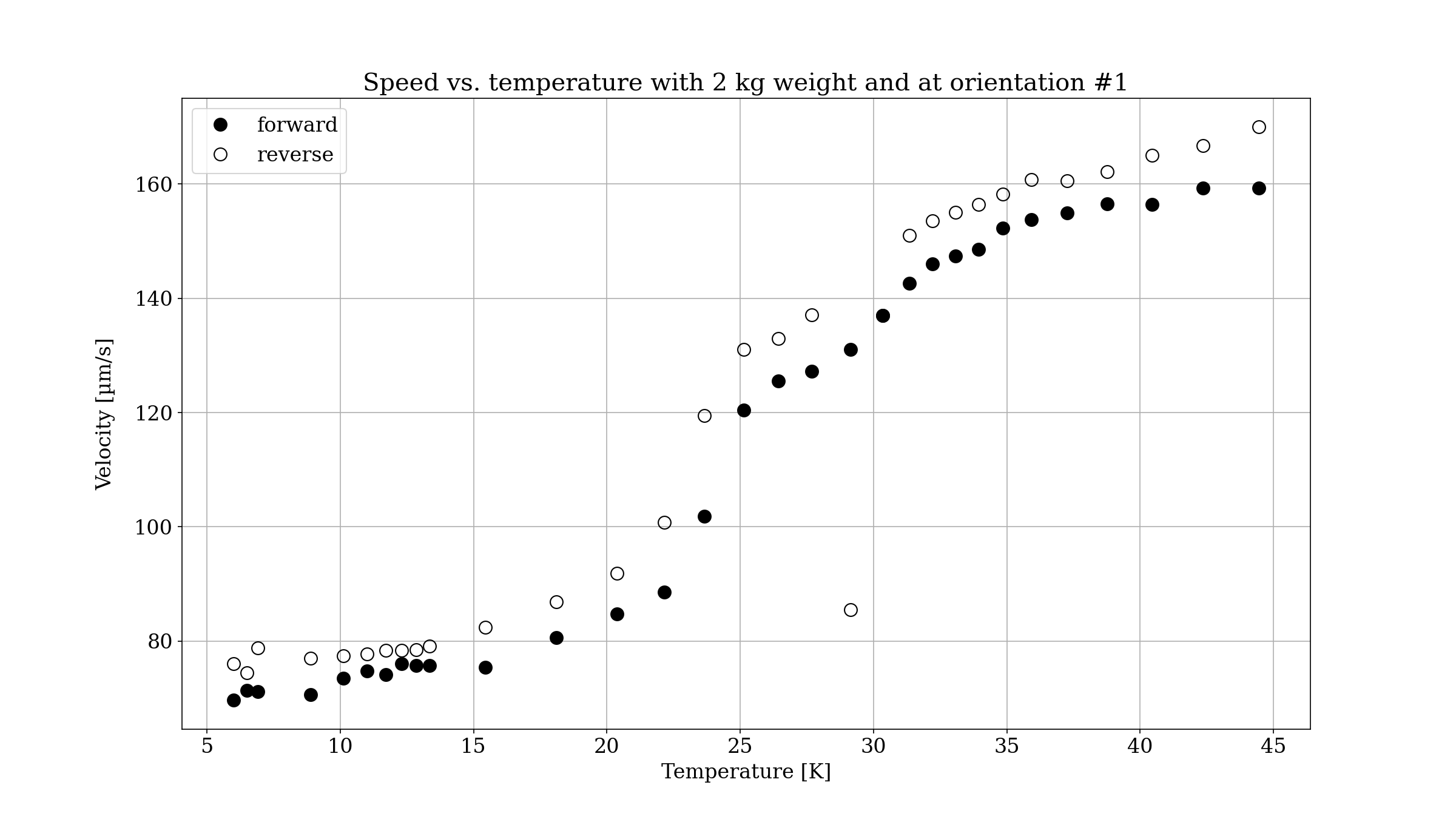}
  \caption{Velocity of the linear stage as function of temperature. Measured with \SI{2}{\kilo\gram} load mounted to the stage carriage and with the piezo actuator moving in orientation \#1. Measurement was performed with the original set-up, and without magnetic field.}
  \label{fig:v_vs_T}
\end{figure}

Stage velocity as function of temperature is shown in figure~\ref{fig:v_vs_T}. At approximately \SI{6}{\kelvin}, the stage (JPE set-up, orientation \#1) was initially moving with a velocity of \SI{70}{\micro\metre\per\second} and \SI{76}{\micro\metre\per\second}, respectively. The cryostat temperature was then slowly increased in a controlled manner. The stage velocity increased, as expected, with increasing actuator temperature (for both movement directions). At lower temperatures, the piezo stroke is, as mentioned above, smaller and so, for a fixed frequency of \SI{50}{\hertz}, the stage velocity is reduced. As the temperature is increased, the piezo stroke becomes larger, resulting in larger velocities, as observed during the experiment. 

\subsection{Measurements at high magnetic fields and in a gaseous helium environment}

\begin{figure}
    \centering
    \includegraphics[width=\textwidth]{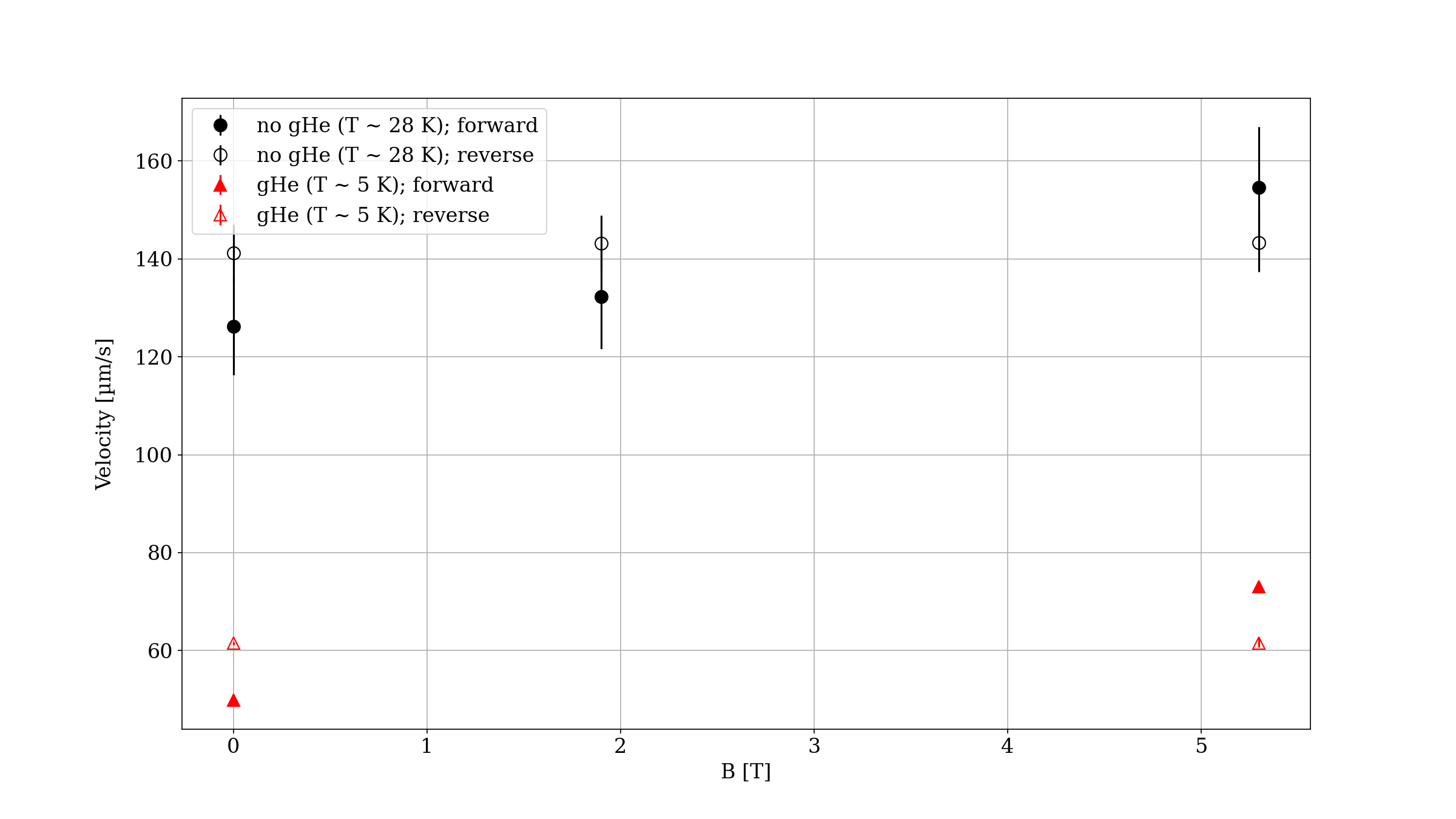}
    \caption{Velocity of the linear stage as a function of the applied magnetic field. Measurements were performed with the ALPS set-up (approximate mechanical load: \SI{1}{\kilo\gram}) and inside the bore of the ALPS magnet with and without a gaseous helium (gHe) atmosphere (pressure between a few \SI{e-2}{\milli\bar} and \SI{e-1}{\milli\bar}) surrounding the set-up. The presence of the gaseous helium atmosphere significantly improves the thermal coupling between the set-up and the cold bore of the magnet. Without the gaseous helium atmosphere, the stage temperature increases during operation, resulting in larger error bars.}
    \label{fig:jpe-speed-alps-magnetic}
\end{figure}

The measurement was repeated with different magnetic field values. The comparison of the stage velocity for three values of the magnetic field (\SIlist{0;1.9;5.3}{\tesla}) is shown in figure~\ref{fig:jpe-speed-alps-magnetic}. Here, both the tests in vacuum ($P < \SI{e-4}{\milli\bar}$) and in a helium gas environment are shown. At $P < \SI{e-4}{\milli\bar}$ and $T = \SI{28}{\kelvin}$ (black markers), in a \SI{1.9}{\tesla} magnetic field, a small increase in velocity from \SI{126}{\micro\metre\per\second} to \SI{132}{\micro\metre\per\second} was observed for the forward direction.  In a \SI{5.3}{\tesla} magnetic field, a significantly larger velocity of \SI{155}{\micro\metre\per\second} was obtained. For the reverse direction, a much smaller increase to \SI{141}{\micro\metre\per\second} and to \SI{143}{\micro\metre\per\second}, respectively,  was observed. The large change of velocity for the forward direction in the presence of a strong magnetic field perpendicular to the direction of movement is still under investigation. The large error bars for the data points without a helium gas atmosphere results from the temperature of the piezo actuator rising during movement cycles of the stage due to the weak thermal linkage of the piezo actuator to the rest of the system.

Helium gas (gHe) was injected gradually into the ALPS magnet, directly before and after ramping up the ALPS magnet to \SI{5.3}{\tesla}. Within \num{10} minutes after injecting gas, the temperature of the set-up dropped from \SI{28}{\kelvin} to approximately \SI{5}{\kelvin}. In the gaseous helium environment, the stage was approximately a factor two slower as expected from the dependence of the stage velocity on the stage temperature (see also figure~\ref{fig:v_vs_T}). This decrease in velocity was observed for all He partial pressures and for both directions of movement. In all previous tests in an environment with $P < \SI{e-4}{\milli\bar}$, an increase of the local temperature of the set-up  during operation could be observed. In the presence of helium gas, this increase in temperature was not observed. This is likely due to the gas efficiently cooling the piezo element. Thus, the actuator does not warm up locally. This keeps the stroke of the piezo element very small, i.e., the linear stage very slow. Additionally, in the presence of helium gas, an increase in velocity in the forward direction at 5.3 T was observed, as already seen in vacuum (Figure~\ref{fig:jpe-speed-alps-magnetic}). 


\section{Conclusions}
In the series of qualification tests and measurements, it could be verified that the demonstrator of the piezo-based positioning stage for the MADMAX booster works in all orientations and mostly independent of the applied mechanical load. The here presented piezo-based linear stage works at room temperature as well as at cryogenic temperatures down to \SI{4.5}{\kelvin}. While the velocity at the lowest temperatures with approximately \SIrange{60}{70}{\micro\metre\per\second} is below the specified value of \SI{100}{\micro\metre\per\second}, it is still sufficient for MADMAX. It could be shown that the linear stage works in high magnetic fields up to \SI{5.3}{\tesla} as well as in a gaseous helium atmosphere. The latter is crucial to improve the thermal linkage between the stage and its surroundings to keep the piezo actuator from warming up during movement. This shows that the actuator developed for MADMAX exploiting the stick-slip principle is well-suited for the disc positioning system of the MADMAX booster.
\section{Outlook}
With the general concept of the piezo-based linear stage being successfully qualified for the MADMAX booster, a simple system, dubbed Project200 (or in short P200) has been built. P200 features three of the piezo stages moving a single \SI{200}{\milli\metre} diameter disc in a small-scale version of the mechanical structure of MADMAX' prototype booster. In addition, Project200 features a cryo-compatible laser interferometer, which allows monitoring the positions of the three piezo actuators, and a dedicated control system. This system allows for the synchronous movement of three piezo actuators using the laser interferometer reading as feedback signal for the positioning of the disc. Tests of the Project200 set-up inside a \SI{1.6}{\tesla} magnetic field and at cryogenic temperatures have been performed and results will be reported in a future publication.

\acknowledgments
The authors would like to thank the DESY-ALPS-II team as well as the DESY-MKS and DESY-MVS teams for the support during the measurements in the ALPS magnet. 

This work is supported by the Deutsche Forschungsgemeinschaft (DFG, German Research Foundation) under Germany’s Excellence Strategy, EXC 2121, Quantum Universe (390833306).



\end{document}